\documentclass[aps,pre,twocolumn,showpacs,groupedaddress]{revtex4}  
\usepackage{graphicx}  
\usepackage{dcolumn}   
\usepackage{bm}        
\usepackage{amssymb}   
\usepackage{float}

\begin{document}



\title{Dark solitons, Breathers and Rogue Wave Solutions of the Coupled Generalized Nonlinear Schr\"{o}dinger Equations}
\author{N. Vishnu Priya, M. Senthilvelan and M. Lakshmanan}
\address{Centre for Nonlinear Dynamics, School of Physics, Bharathidasan University, Tiruchirappalli - 620 024, Tamil Nadu, India.}

\begin{abstract}
We construct dark-dark solitons, general breather (GB), Akhmediev breather (AB), Ma soliton (MS) and rogue wave (RW) solutions of a coupled generalized nonlinear Schr\"{o}dinger  (CGNLS) equation.  While the dark-dark solitons are captured in the defocusing regime of CGNLS system, the other solutions, namely GB, AB, MS and RW, are  identified in the focusing regime.  We also analyze the structures of GB, AB, MS and RW profiles with respect to the four-wave mixing parameter.  We show that when we increase the value of the real part of the four wave mixing parameter, the number of peaks in the breather profile increases and the width of each peak gets shrunk.  Interestingly the direction of this profile also changes due to this change.  As far as the RW profile is concerned the width of the peak becomes very thin when we increase the value of this parameter.  Further, we consider RW solution as the starting point and derive AB, MS and GB in the reverse direction and show that the solutions obtained in both the directions match with each other.  In the course of the reverse analysis we also demonstrate how to capture the RW solutions directly from AB and MS.  
\end{abstract}

\pacs{02.30.Ik, 42.65.-k, 47.20.Ky, 05.45.Yv}
\maketitle


\section{Introduction} 
In this paper we construct dark-dark soliton, Akmediev breather, Ma soliton, general breather and rogue wave (RW) solution of an integrable two coupled generalized nonlinear Schr\"{o}dinger equation (CGNLS), namely \cite{Wang}
\begin{eqnarray}
\label{cnls01}
\nonumber ip_{t}+p_{xx}+2(a|p|^2+c|q|^2+bpq^*+b^*qp^*)p&=&0,\\
iq_{t}+q_{xx}+2(a|p|^2+c|q|^2+bpq^*+b^*qp^*)q&=&0,
\label{cnls01}
\end{eqnarray}
where $p$ and $q$ are slowly varying pulse envelopes and $a$ and $c$ are real constants corresponding to self phase modulation and cross phase modulation effects, respectively.  Here $b$ is a complex constant corresponding to four wave mixing and $*$ denotes complex conjugation.  Besides self phase and cross phase modulation effects, four wave mixing effects have also been included in (\ref{cnls01}).  When $a=c$ and $b=0$ the above equation reduces to the Manakov system \cite{Manakov}.  When $a=-c$ and $b=0$ it reduces to the mixed coupled nonlinear Schr\"{o}dinger equation \cite{viji}.  The nonlinear system (\ref{cnls01}) has also been proved to be completely integrable for arbitrary values of the system parameters $a$, $b$ and $c$ through Weiss-Tabor-Carnevale (WTC) algorithm \cite{Penj}.  N-bright soliton formula for the system (\ref{cnls01}) is constructed in Ref. \cite{Wang}.  Except for the bright N-soliton solution no other solution is known for this system (\ref{cnls01}).   The aim of this paper is to derive dark solitons and certain rational solutions including RW solutions for the nonlinear evolutionary equation (\ref{cnls01}) and analyze how the solution profiles vary with respect to the four wave mixing parameter $b$.

\par RWs (or freak waves, extreme waves) are large amplitude ocean waves which are capable of making disastrous effects to oil tanks and cruise ships, which were first considered as mysterious until recorded for the first time by scientific measurements.  Consequently the characteristics of the RWs have been analyzed in various perspectives, see for example \cite{Pelin} and references therein.  Recently RW solutions have been identified in different areas of physics - to name a few we cite the situations in fiber optics \cite{Solli}, water tank experiments \cite{watertank}, in plasma physics \cite{Plasma}, Bose-Einstein condensates (BEC) \cite{BEC} and so on.  In particular optical RWs arise due to supercontinuum generation \cite{Solli} in optical fibers and they have many potential applications to generate stable, coherent light sources, and to produce high contrast optical switches \cite{Turitsyn}.  Mathematically, the first and simplest RW solution was reported for the nonlinear Schr\"{o}dinger equation by Peregrine \cite{Peregrine}.  This solution approaches a non-zero constant background as time goes to infinity but develops a localized hump with peak amplitude three times the constant background in the intermediate times.  Very recently higher order RWs were also reported for certain nonlinear evolution equations, see for example Refs. \cite{{Akhmediev2}, {Guo}, {zhaqilao}, {Porsezian}, {3coupled}}.  It has been shown that these higher order RWs could reach higher peak amplitudes or exhibit multiple intensity peaks at different spatial locations and times.  From the mathematical point of view these RWs can be a limiting case of time periodic breather or Ma soliton (MS) and space periodic breather or Akhmediev breather (AB) \cite{note}.  These breathers arise due to the effect of modulation instability, which is a characteristic feature of a wide class of nonlinear dispersive systems, associated with dynamical growth and evolution of periodic perturbation on a continuous wave background  \cite{MI}.  This dynamics is closely related to the celebrated Fermi-Pasta-Ulam recurrence phenomenon, as well as it has potential applications in ultrashort pulse train generation and parametric amplification.  One way of obtaining RW solution or Peregrine soliton for a given nonlinear partial differential equation is to construct a breather solution, either AB or MS from the which the RW solution can be deduced in an appropriate limit. 

\par Even though RW solutions have been constructed for nonlinear Schr\"{o}dinger equation, derivative nonlinear Schr\"{o}dinger equation and its generalizations and certain multi-component analogues \cite{{Kalla}, {Zhai}, {Ling}, {Degasperis}, {Vishnu}}, to our knowledge, RWs in coupled NLS equation with higher order effects such as four wave mixing effects have not been studied so far.  Four wave mixing is a basic nonlinear phenomenon having fundamental relevance and practical applications particularly in nonlinear optics \cite{Agrawal}, optical processing \cite{Pepper}, phase conjugate optics \cite{Yariv}, real time holography \cite{Gerritsen} and measurement of atomic energy structures and decay rates \cite{{Bjorklund},{Yajima}}.  Motivated by these considerations and importance of breathers and RWs, it is an interesting problem to find how these four wave mixing terms will affect the breather and RW solutions in an associated nonlinear system.  Hence, in the present work we construct dark-dark soliton, AB, MS, GB and RW for (\ref{cnls01}).  Conventionally these solutions are constructed through Darboux transformation method.  Differing from this approach we derive all these solutions through Hirota's bilinearization method.  To begin with, using Hirota's method, we construct dark-dark soliton solution for the defocusing CGNLS Eq. (\ref{cnls01}) with the parameters $a$ and $c$ replaced by $-a$ and $-c$ ($a>0$, $c>0$) respectively.  To construct GB, we bilinearize the focusing CGNLS equation ($a>0$, $c>0$ in Eq. (\ref{cnls01})) and construct two soliton solution.  By appropriately restricting the parameters which appear in the phase factor we capture the GB, AB, MS and RWs.  We then analyze mathematically how these solution profiles change with respect to four wave mixing terms.  As far as the GB profile is concerned our result shows that when we increase the value of the real value of the four wave mixing parameter not only the number of peaks increases but also the direction of the underlying profile gets altered.  As we expect in the AB and MS profiles when we increase the  value of $Re$ $b$ only the number of peaks increases.  In the case of RW solution the profile becomes thinner and thinner when we increase the value of $Re$ $b$.  Our studies also reveal that the value of $Im$ $b$ does not change either the number of pulses or the direction of these profiles.

\par We also analyze the reverse problem of constructing a AB or MS or GB from a RW solution.  To derive them we rewrite the RW solution in a factorized form and generalize this factorized form in an imbricate series expression \cite{Toda} with certain unknown parameters in this series.  We then find these unknown parameters by substituting them in Eq. (\ref{cnls01}) and solving the resultant equations.  With three different forms of the imbricate series we derive the AB, MS and GB solutions from the RW solution of (\ref{cnls01}).  In addition to the above, we demonstrate how to derive RW solution directly from AB or MS.  To demonstrate this first we point out that these two profiles strongly depend on a critical parameter.  And by making a Taylor expansion of these solutions at the critical parameter value we obtain the RW solution.  We also illustrate the isolation of RW from AB and MS pictorially. 

\par The plan of the paper is follows.  In Sec.\ref{Ds} we derive the explicit form of dark-dark soliton solutions of the defocusing CGNLS equation.  In Sec. \ref{GB} we construct the explicit form of the GB solution of the CGNLS system (\ref{cnls01}) for the focusing case through Hirota's bilinearization method.  We then explain the method of deriving AB, MS and RW solutions from the GB solution.  In Sec. \ref{RWtoAB}, we discuss the method of constructing the AB solution from the RW solution.  In Sec. \ref{RWtoMS}, we demonstrate the construction of MS from RW.  In Sec. \ref{RWtoGB}, we formulate the imbricate series form for the RWs with certain unknown arbitrary functions in it and then compare this expression with the one derived from the GB in the same way.  The comparison provides exact expressions for the unknown arbitrary functions which appear in the imbricate series of the RW.  In this way we establish a method of constructing GBs from RW.  Finally, in Sec. \ref{conclusions} we present our conclusions.  

\section{Defocusing Case: Dark solitons}
\label{Ds}
To start with we construct dark-dark solitons for (\ref{cnls01}) for the defocusing case by using Hirota's method.  Using the transformation $p=\frac{g}{f}$ and $q=\frac{h}{f}$, the defocusing CGNLS system
\begin{eqnarray}
\label{cnls02}
\nonumber ip_{t}+p_{xx}+2(-a|p|^2-c|q|^2+bpq^*+b^*qp^*)p&=&0,\\
iq_{t}+q_{xx}+2(-a|p|^2-c|q|^2+bpq^*+b^*qp^*)q&=&0,\nonumber \\a,c&>&0,
\label{pct1}
\end{eqnarray}
can be bilinearized as 
\begin{eqnarray}
(iD_t+D_x^2-\lambda)g.f=0, \nonumber\\
(iD_t+D_x^2-\lambda)h.f=0, \nonumber\\
(D_x^2-\lambda)f.f=-2(a|g|^2+c|h|^2-bgh^*-b^*hg^*)
\label{ds3}   
\end{eqnarray}
in which $\lambda$ is a constant to be determined.  In the above, $D_t$ and $D_x$ are Hirota's bilinear operators.  To construct a dark one-soliton solution of (\ref{cnls02}), we assume $g=g_0(1+\chi g_1)$, $h=h_0(1+\chi h_1)$ and $f=1+\chi f_1$, where $g_i's$ and $h_i's$, $i=0,1$, are complex functions of $x$ and $t$ and $f_i's$ are real functions.  Substituting these forms into (\ref{ds3}) and then collecting the coefficients of $\chi^{0} $, we get  
\begin{figure*}
\includegraphics[width=0.7\linewidth]{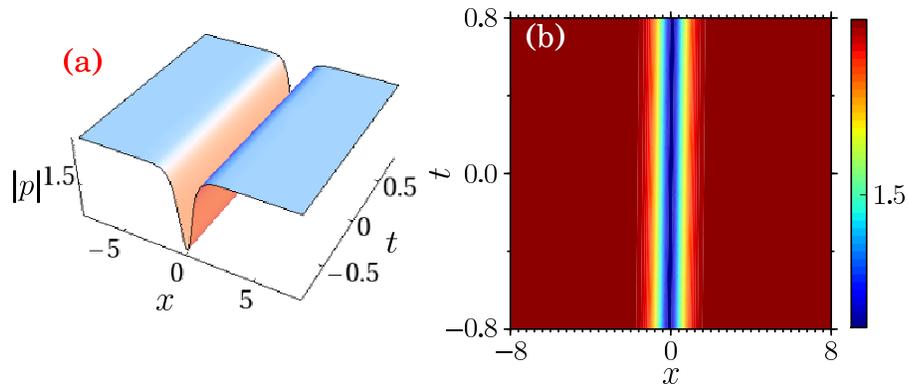}
\caption{(a) Dark-dark soliton profile of $p$ for the parameter values $\tau_1=0.8$, $\tau_2=1.5$, $a=1$, $c=.5$, $k=0.5$, $\Omega_1=1$ and $P_1=0.5$.  (b) Corresponding contour plot.  Similar profile occurs for $q$ also (not shown here)(Color online).}
\end{figure*}
\begin{eqnarray}
(iD_t+D_x^2-\lambda)g_0.1=0, \nonumber\\ 
(iD_t+D_x^2-\lambda)h_0.1=0, \nonumber\\ 
2(ag_0g_0^*+ch_0h_0^*-bg_0h_0^*-b^*h_0g_0^*)=\lambda.
\label{ds5}
\end{eqnarray}  
\par Eq. (\ref{ds5}) admits the following solution,
\begin{eqnarray}
g_0=\tau _1\exp(i\psi_1), \;\;    h_0=\tau _2\exp(i\psi _2),
\label{ds6}
\end{eqnarray}
where 
\begin{eqnarray}
\psi_i=k_ix-(\lambda+k_i^2)t+\psi_i^{(0)},  \;\;  i=1,2,
\label{ds61}
\end{eqnarray}
in which $k_i$, $\psi_i^{(0)}$ and $\tau _i$  are real constants.
\par To explore dark-dark solitons, we look for solutions with the following large-distance asymptotics, that is $p\rightarrow\tau _1\exp(i\psi _1)$ and $q\rightarrow\tau _2\exp(i\psi_2)$.  Substituting these forms into the CGNLS system (\ref{cnls02}), we observe that the presence of four wave mixing terms in Eq. (\ref{cnls02}) require us to set the constants $k_1$ and $k_2$ to be equal, that is $k_1=k_2\equiv k$, in order to get a unique dispersion relation.  In other words the phase factors $\psi_1$ and $\psi_2$ now become one and the same, that is $\psi_1=\psi_2\equiv\psi$.  With this restriction the last expression in (\ref{ds5}) yields the constraint
\begin{eqnarray}
a\tau_1^2+c\tau_2^2-b\tau_1\tau_2-b^*\tau_1\tau_2=\lambda/2.
\label{t1}
\end{eqnarray}
\par Taking into account the forms (\ref{ds6}) and the usual Hirota identities, the coefficients of $\chi$ lead to 
\begin{eqnarray}
(iD_t+2ikD_x+D_x^2)(1.f_1+g_1.1)&=&0,  \nonumber \\ 
(iD_t+2ikD_x+D_x^2)(1.f_1+h_1.1)&=&0,  \nonumber\\
(D_x^2-\lambda)(1.f_1+f_1.1)+2(a\tau_1^2(g_1+g_1^*)&&\nonumber\\+c\tau_2^2(h_1+h_1^*)-b\tau_1\tau_2-b^*\tau_1\tau_2)&=&0.
\label{ds7}
\end{eqnarray}
Equation (\ref{ds7}) admits the following solutions,
\begin{eqnarray}
g_1=h_1=Zexp(\zeta), \;\; f_1=exp(\zeta),
\label{ds8}
\end{eqnarray}
where
\begin{eqnarray}
\zeta=\Omega_1t-P_1x+\zeta^{(0)}
\label{ds9}
\end{eqnarray}
in which $P_1$, $\Omega_1$ and $\zeta^{(0)}$ are real constants and $Z$ is a complex constant.  These constants are interconnected by the relations
$Z=\frac{-(P_1^2-i(\Omega_1-2l_1P_1))^2}{P_1^4+(\Omega_1-2l_1P_1)^2}$ and
$\frac{a\tau_1^2+c\tau_2^2-(b+b^*)\tau_1\tau_2}{P_1^4+(\Omega_1-2l_1P_1)^2}=\frac{1}{4P_1^2}$.  It can easily be checked that $|Z|^2=1$.  Substituting the obtained forms of $g_1$, $h_1$ and $f_1$ in the expressions $p=\frac{g}{f}$ and $q=\frac{h}{f}$,  the dark one-soliton solution can be identified in the following form (after suitably absorbing $\chi$), that is
\begin{eqnarray}
p&=&\frac{\tau_1}{2}\exp(i\psi)[(1+Z)-(1-Z)\tanh(\zeta_1/2)],\nonumber\\
q&=&\frac{\tau_2}{2}\exp(i\psi)[(1+Z)-(1-Z)\tanh(\zeta_1/2)].
\label{ds14}
\end{eqnarray}
The obtained solution is plotted in Fig.1 which shows a regular behavior of a dark soliton which has a localized intensity dip.  One may also observe that as $x=\pm\infty$ the solution attains a plane wave form.
\par For constructing two dark-dark soliton solution we assume, 
\begin{eqnarray}
g=g_0(1+\chi g_1+\chi^2g_2),\; h=h_0(1+\chi h_1+\chi^2h_2),\nonumber\\ {\text{and}}\; f=1+\chi f_1+\chi^2f_2,
\label{2dd}
\end{eqnarray}
where $g_0$ and $h_0$ are obtained here as in Eqs. (\ref{ds6}) and (\ref{ds61}).  Substituting Eq. (\ref{2dd}) in Hirota bilinear equation (\ref{ds3}) and equating the various coefficients of the powers of $\chi$ to zero we will get a set of equations.  By solving these equations appropriately we can obtain exact expressions for $g$, $h$ and $f$.  Substituting these expressions in the transformation $p=\frac{g}{f}$ and $q=\frac{h}{f}$ we can derive two dark-dark soliton solution.  By assuming, $g=g_0(1+\chi g_1+...+\chi^Ng_N)$, $h=h_0(1+\chi h_1+...+\chi^Nh_N)$, $f=1+\chi f_1+...+\chi^Nf_N$ and proceeding in the same way as discussed above one can derive N- dark-dark soliton solutions.
\par We wish to note here the fact that dark solitons are more robust than bright solitons due to their background energy.  Potential applications of dark solitons have been proposed in many contexts such as optical fibers \cite{Kivshar}, Bose-Einstein condensates \cite{becapp}, plasma physics and so on.  Our results on dark solitons for CGNLS system (\ref{pct1}) which includes four wave mixing effects can have potential realization in optical wave propagation in birefringent optical fibers, and matter wave propagation in Bose-Einstein condensates.
\section{Focusing case: General Breathers} 
\label{GB}
Now we turn our attention to the focusing case, Eq. (\ref{cnls01}) with $a>0$, $c>0$.  To derive the breather solution we seek a two periodic soliton solution to the focusing CGNLS equation (\ref{cnls01}) with the boundary conditions $|q_i|^2 \to \tau_i^2$, $i=1,2,$ as $x \to \pm\infty$, where $\tau_1$ and $\tau_2$ are real constants.  We again bilinearize Eq. (\ref{cnls01}), with $a>0$, $c>0$.  The resultant bilinearized form will be of the same form as given in Eq. (\ref{ds5}) with the only difference in the signs in front of the constants $a$ and $c$ which now become positive.

\par As our aim is to obtain a two periodic soliton solution we terminate the expansion at quadratic powers in $\chi$, that is $g=g_0(1+\chi g_1+\chi^2g_2)$, $h=h_0(1+\chi h_1+\chi^2h_2)$ and $f=(1+\chi f_1+\chi^2f_2)$.  Substituting these forms into the bilinearized equations and solving the resultant equations we obtain the exact expressions for $g$, $h$ and $f$.  The resultant two soliton solution emerges in the form
\begin{eqnarray}
p&=&\tau_1e^{i(kx-\omega t)}\nonumber\\&&\times \left(\frac{1+e^{\eta_1+2i\phi_1}+e^{\eta_2+2i\phi_2}+\vartheta e^{\eta_1+\eta_2+2i\phi_1+2i\phi_2}}{1+e^{\eta_1}+e^{\eta_2}+\vartheta e^{\eta_1+\eta_2}}\right), \nonumber \\
q&=&\tau_2e^{i(kx-\omega t)}\nonumber\\&&\times \left(\frac{1+e^{\eta_1+2i\phi_1}+e^{\eta_2+2i\phi_2}+\vartheta e^{\eta_1+\eta_2+2i\phi_1+2i\phi_2}}{1+e^{\eta_1}+e^{\eta_2}+\vartheta e^{\eta_1+\eta_2}}\right),\nonumber\\
\label{pct3}
\end{eqnarray}
where $\eta_j=p_jx-\Omega_jt+\eta_j^0, \ j=1,2$.  In the above $p_j$, $\Omega_j$, $\eta_j^0$ and $\phi_j$, $j=1,2$, are complex parameters and
\begin{eqnarray}
\omega&=&k^2-2(a\tau_1^2+c\tau_2^2+(b+b^*)\tau_1\tau_2), \nonumber\\ p_j&=&2i\sqrt{(a\tau_1^2+c\tau_2^2+(b+b^*)\tau_1\tau_2)}\sin\phi_j,\nonumber\\
\Omega_j&=&2k_jp_j-p_j^2\cot\phi_j, \;\vartheta=\left(\frac{\sin\frac{1}{2}(\phi_1-\phi_2)}{\sin\frac{1}{2}(\phi_1+\phi_2)}\right)^2,   \nonumber\\  \phi_j&=&\phi_{jR}+i\phi_{jI}, \; j=1,2.
\label{pct4}
\end{eqnarray}

\par In the following we show that by appropriately restricting the complex parameters one can rewrite the above breather expression into Akhmediev and Ma breathers.  To obtain the breather solution from the above two soliton solution we take $\eta_1=\eta_2^*\equiv \eta$ and $\phi_2=\phi_1^*\pm \pi$.  Substituting these two restrictions in (\ref{pct3}) and considering $\eta=\eta_{R}+i\eta_{I}$ and  $\phi_1=\phi_{R}+i\phi_{I}$, the exponential functions appearing in (\ref{pct3}) can be rewritten in terms of trigonometric and hyperbolic functions.  The resultant action leads us to                                                                                    
\begin{eqnarray}
p&=&\tau_1\cos2\phi_R e^{i(\theta+2\phi_R)}\bigg[1+\frac{1}{\sqrt{\vartheta}\cosh(\eta_R+\sigma)+\cos\eta_I}\nonumber\\&&\times\bigg(\bigg(\frac{\cosh2\phi_I}{\cos2\phi_R}-1\bigg)\cos\eta_I+i\big(\sqrt{\vartheta}\tan2\phi_R\nonumber\\&&\times\sinh(\eta_R+\sigma)-\frac{\sinh2\phi_I}{\cos2\phi_R}\sin\eta_I\big)\bigg)\bigg],\nonumber\\
q&=&\tau_2\cos2\phi_R e^{i(\theta+2\phi_R)}\bigg[1+\frac{1}{\sqrt{\vartheta}\cosh(\eta_R+\sigma)+\cos\eta_I}\nonumber\\&&\times\bigg(\bigg(\frac{\cosh2\phi_I}{\cos2\phi_R}-1\bigg)\cos\eta_I+i\big(\sqrt{\vartheta}\tan2\phi_R\nonumber\\&&\times\sinh(\eta_R+\sigma)-\frac{\sinh2\phi_I}{\cos2\phi_R}\sin\eta_I\big)\bigg)\bigg],
\label{pct5}
\end{eqnarray}
where $\eta_R=p_Rx-\Omega_Rt+\eta_R^0$,  $\eta_I=\Omega_It-p_Ix+\eta_I^0$, $p_1=p_R+ip_I$, $\Omega_1=\Omega_R+i\Omega_I$, $\eta_R^0$, $\eta_I^0$ and $\sigma$ are constants.  The exact forms of $p_R$, $\Omega_R$, $p_I$, $\Omega_I$ and $\vartheta$ are given below:
\begin{eqnarray}
p_R&=&-2\sqrt{(a\tau_1^2+c\tau_2^2+(b+b^*)\tau_1\tau_2)}\cos\phi_R\sinh\phi_I, \nonumber\\ p_I&=&2\sqrt{(a\tau_1^2+c\tau_2^2+(b+b^*)\tau_1\tau_2)}\sin\phi_R\cosh\phi_I,\nonumber\\
\Omega_R&=&2kp_R-\frac{(p_R^2-p_I^2)\sin 2\phi_R+2p_Rp_I\sinh 2\phi_I}{\cosh 2\phi_I-\cos 2\phi_R}, \nonumber\\ \Omega_I&=&2kp_I+\frac{(p_R^2-p_I^2)\sinh 2\phi_I-2p_Rp_I\sin 2\phi_R}{\cosh 2\phi_I-\cos 2\phi_R},\nonumber\\
\vartheta&=&\frac{\cosh^2\phi_I}{\cos^2\phi_R}.
\label{pc5}
\end{eqnarray}
\begin{figure*}
\includegraphics[width=0.8\linewidth]{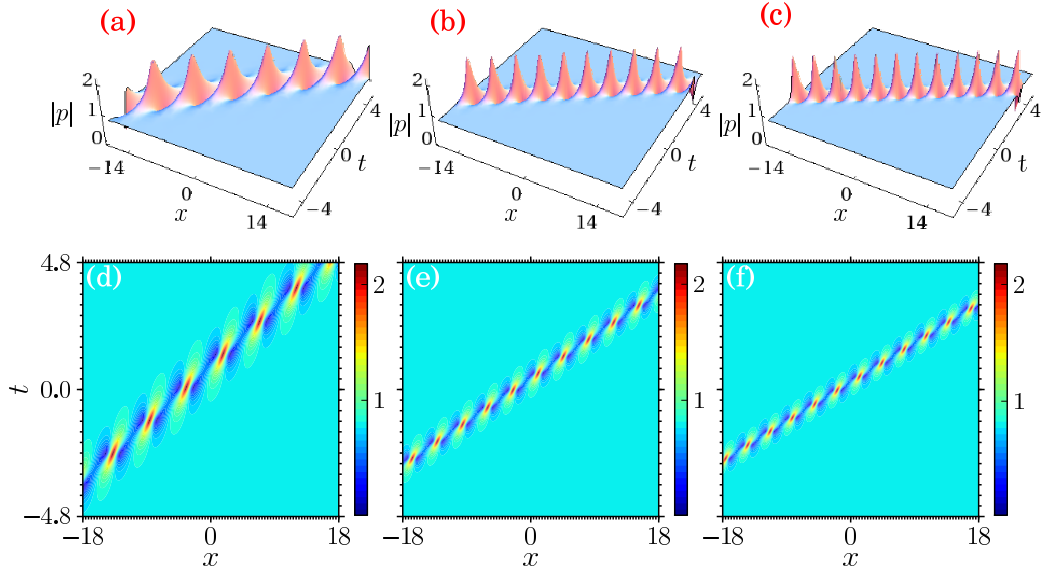}
\caption{General breather profiles of $p$ for the parameter values $\tau_1 = 0.8$, $\tau_2=0.5$, $a=0.5$, $c=0.5$, $\phi_R = 0.7$, $\phi_I = -0.5$, $\eta_I^o=1.5$, $\eta_R^o=1.2$, $k=0.31$ with different four wave mixing parameter values (a) $b=0.5+i$, (b) $b=1.5+i$ (c) $b=2+i$.  Figs. (d)-(f) are the corresponding contour plots of (a)-(c) respectively.  Similar profile occurs for $q$ also (not shown here) (Color online).}
\end{figure*}

\par Consequently the expressions (\ref{pct5}) with (\ref{pc5}) constitute the GB solution of the CGNLS Eq. (\ref{cnls01}).  Figs. 2 illustrate the behavior of this breather solution, which is periodic both in space and time.  As our aim is to analyze the effect of the four wave mixing parameter $b$, we vary this parameter and study the outcome.  For example, for $b=0.5+i$ we get a plot as shown in Figs. 2(a) and 2(d) which contain six full peaks during the time interval $t = -4.8 \; to \; 4.8$ and space interval $x=-18\; to \; 18$.  If we now change the value of $b$ to $b=1.5+i$ so that $Re$ $b$ is increased from 0.5 to 1.5, we observe that the width of the peak decreases and we get ten full peaks for the same space and time intervals which is demonstrated in Figs. 2(b) and 2(e).  When we increase further the value of $Re$ $b$, we observe that the number of the peaks gradually increases whereas the width of the pulses gets shrunk which is demonstrated in Figs. 2(c) and 2(f).  We can confirm our results analytically.  To do so, we identify two successive maximum points of the GB solution and show that the distance between these two maximum points is proportional to the real part of the four wave mixing parameter $b$, namely Re $b$.
\par To find a maximum point of GB solution, we need to find at which values of $x$ and $t$, $\frac{\partial |p|}{\partial x}$ and $\frac{\partial |p|}{\partial t}$ become zero.  For this purpose we rewrite the Eq. (\ref{pct5}) as
\begin{eqnarray}
|p|&=&\frac{\tau_1}{\sqrt{a}\cosh(\eta+\sigma)+\cos\eta_I}\big((\sqrt{a}\cos2\phi_R\cosh(\eta_R+\sigma)\nonumber\\&&+\cosh2\phi_I\cos\eta_I)^2+(\sqrt{a}\sin2\phi_R\sinh(\eta_R+\sigma)\nonumber\\&&-\sinh2\phi\sin\eta_I)^2\big)^{1/2}.
\label{modp}
\end{eqnarray} 
Now we differentiate Eq. (\ref{modp}) with respect to $x$ and $t$ we obtain the expressions for $\frac{\partial |p|}{\partial x}$ and $\frac{\partial |p|}{\partial t}$ which are given in Appendix.  From these expressions, we observe that at the points
\begin{eqnarray}
(x,t)=\left(\frac{\pm n\pi}{p_I}+\frac{\Omega_I(\pm n\pi p_R+p_I\sigma)}{p_I(p_I\Omega_R-p_R\Omega_I)},\frac{\pm n\pi p_R+p_I\sigma}{p_I\Omega_R-p_R\Omega_I}\right),\nonumber\\n=0,1,2,\cdots,\nonumber\\
\label{xt1}
\end{eqnarray}
$\frac{\partial |p|}{\partial x}$ and $\frac{\partial |p|}{\partial t}$ become zero.  So these are the stationary points of the GB solution.  Now we need to classify whether these points are saddles or extreme points.  For this purpose we need to do the following second derivative test.
\begin{figure}
\includegraphics[width=0.7\linewidth]{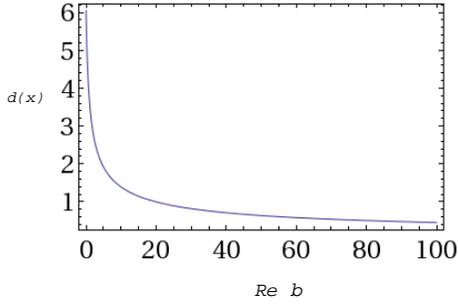}
\caption{Distance plot for the parameter values $\tau_1 = 0.8$, $\tau_2=0.5$, $a=0.5$, $c=0.5$, $\phi_R = 0.7$, $\phi_I = -0.5$, $\eta_I^o=1.5$, $\eta_R^o=1.2$, $k=0.31$ (Color online).}
\end{figure}
\par Let ($\alpha$, $\beta$) be a stationary point, so that $|p|_x=0$ and $|p|_t=0$ at ($\alpha$, $\beta$).  Then:\\
(i) If $|p|_{xx}|p|_{tt}-|p|^2_{xt}<0$ at ($\alpha$, $\beta$), then ($\alpha$, $\beta$) is a saddle point.\\
(ii) If $|p|_{xx}|p|_{tt}-|p|^2_{xt}>0$ at ($\alpha$, $\beta$), then ($\alpha$, $\beta$) is a maximum point when $|p|_{xx}<0$ and $|p|_{tt}<0$, and ($\alpha$, $\beta$) is a minimum point when $|p|_{xx}>0$ and $|p|_{tt}>0$.
Now applying this procedure on the point with $n=0$ in (\ref{xt1}) we observe that $|p|_{xx}|p|_{tt}-|p|^2_{xt}<0$ so that it is a saddle point. 
\par By following the above procedure to $n=1$ in (\ref{xt1}) we find that at
\begin{eqnarray}
(x,t)=\left(\frac{\pi}{p_I}+\frac{\Omega_I}{p_I}\left(\frac{p_R\pi+p_I\sigma}{p_I\Omega_R-p_R\Omega_I}\right),\frac{p_R\pi+p_I\sigma}{p_I\Omega_R-p_R\Omega_I}\right),\nonumber\\
\end{eqnarray}
the GB solution has a maximum.  The next maximum occurs at
\begin{eqnarray}
(x,t)=\left(\frac{3\pi}{p_I}+\frac{\Omega_I}{p_I}\left(\frac{3p_R\pi+p_I\sigma}{p_I\Omega_R-p_R\Omega_I}\right),\frac{3p_R\pi+p_I\sigma}{p_I\Omega_R-p_R\Omega_I}\right).\nonumber\\
\end{eqnarray}
The distance between these two maximum points along the $x$ direction is
\begin{eqnarray}
d(x)=\frac{2\pi}{p_I}+\frac{\Omega_I}{p_I}\left(\frac{2p_R\pi}{p_I\Omega_R-p_R\Omega_I}\right)
\label{dx1}
\end{eqnarray}
while the time interval between the two peaks is
\begin{eqnarray}
d(t)=\frac{2p_R\pi}{p_I\Omega_R-p_R\Omega_I},
\label{dt1}
\end{eqnarray}
where $p_R$, $p_I$, $\Omega_R$ and $\Omega_I$ have the forms
\begin{eqnarray}
p_R&=&-2\sqrt{(a\tau_1^2+c\tau_2^2+(b+b^*)\tau_1\tau_2)}\cos\phi_R\sinh\phi_I, \nonumber\\
p_I&=&2\sqrt{(a\tau_1^2+c\tau_2^2+(b+b^*)\tau_1\tau_2)}\sin\phi_R\cosh\phi_I,\nonumber\\
\Omega_R&=&\frac{(p_R^2-p_I^2)\sin 2\phi_R+2p_Rp_I\sinh 2\phi_I}{\cosh 2\phi_I-\cos 2\phi_R}, \nonumber\\ \Omega_I&=&-\frac{(p_R^2-p_I^2)\sinh 2\phi_I-2p_Rp_I\sin 2\phi_R}{\cosh 2\phi_I-\cos 2\phi_R}.
\label{pc51}
\end{eqnarray}
From the expressions (\ref{dx1}) and (\ref{dt1}), it is clear that the distance and time interval between two adjacent peaks are inversely proportional to the square root of $Re$ $b$ as shown in Fig. 3 for the case of distance.  Hence, if we increase the value of $Re$ $b$, the distance between the peaks decreases and so the number of the peaks gets increased.
\par In addition to the above, we also noticed that when we increase the value of $Re$ $b$, the direction of the profile also gets changed which can be easily verified from Figs. 2.  To prove this analytically we find the angle $\varphi$ between the $x$ and $t$ axes takes the form
\begin{eqnarray}
\cos\varphi=\frac{\pi}{p_I}\left(\frac{p_I\Omega_R-p_R\Omega_I}{p_R\pi+p_I\sigma}\right)+\frac{\Omega_I}{p_I},
\label{angle}
\end{eqnarray}
where $p_R$, $p_I$, $\Omega_R$ and $\Omega_I$ take the forms as in Eqs. (\ref{pc51}).  From Eq. (\ref{angle}), one can see that the propagation direction of the profile also depends on the real part of the four wave mixing parameter $b$.

\subsection{Akhmediev breather}
\begin{figure*}
\includegraphics[width=0.8\linewidth]{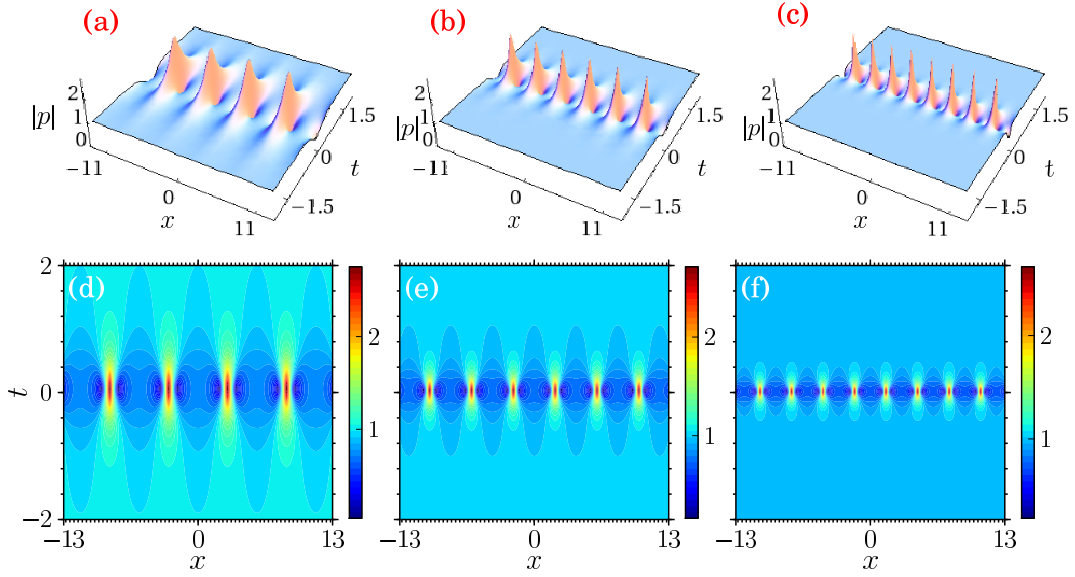}
\caption{Akhmediev breather profile of $p$ for the parameter values  $\tau_1 = 1$, $\tau_2=0.5$, $a=1$, $c=0.5$, $\phi_R = 0.5$, $\eta_I^o=1.5$, $\eta_R^o=1.2$, $k=0$ with different four wave mixing parameter values (a)$b=0.2+i$, (b) $b=1.5+i$, (c) $b=3.5+i$.  Figs. (d)-(f) are the corresponding contour plots of (a)-(c) respectively.  Similar profile occurs for $q$ also (not shown here)(Color online).}
\end{figure*}
From the GB solution we can derive AB, MS and RW solutions by restricting the parameters $\phi_R$ and $\phi_I$ suitably.  In the following, we report the explicit forms of these solutions.  
To derive the AB solution, we fix the wave number to be pure imaginary, that is $\phi_R\neq 0$ and $\phi_I=0$ in (\ref{pct5}).  With this choice Eq. (\ref{pct5}) becomes
\begin{eqnarray}
p&=&\tau_1\cos(2\phi_R)e^{i(\theta+2\phi_R)}\left(1+\frac{A}{B}\right),\nonumber\\
q&=&\tau_2\cos(2\phi_R)e^{i(\theta+2\phi_R)}\left(1+\frac{A}{B}\right),\nonumber\\
A&=&(\frac{1}{\cos(2\phi_R)}-1)\cos\eta_I+i\sqrt{\vartheta}\tan(2\phi_R)\sinh(\eta_R+\sigma),\nonumber\\
B&=&\sqrt{\vartheta}\cosh(\eta_R+\sigma)+\cos\eta_I.
\label{pct7}
\end{eqnarray} 

Here $\eta_R=-\Omega_Rt+\eta_R^0$, $\eta_I=p_Ix+\eta_I^0$, $\theta=-\omega t$, $\vartheta=\frac{1}{\cos^2\phi_R}$.  We choose $k=0$ for covenience.

The AB solution (\ref{pct7}) is periodic in $x$ and localized in $t$.  This spatially periodic breather is nothing but the AB solution which occurs due to the modulational instability process \cite{MI}.  Here also we analyze the effect of $b$ mathematically as discussed in the previous section.  We find a maximum point of AB solution at
\begin{eqnarray}
(x,t)=\left(\frac{\pi}{p_I},\frac{\sigma}{\Omega_I}\right).
\end{eqnarray}
The next maximum point occurs at
\begin{eqnarray}
(x,t)=\left(\frac{3\pi}{p_I},\frac{\sigma}{\Omega_I}\right).
\end{eqnarray}
The distance between these two maximum points is 
\begin{eqnarray}
d(x)=\frac{2\pi}{p_I}=\frac{2\pi}{2\sqrt{(a\tau_1^2+c\tau_2^2+(b+b^*)\tau_1\tau_2)}\sin\phi_R\cosh\phi_I}.\nonumber\\
\label{dx2}
\end{eqnarray}
From Eq. (\ref{dx2}) we can observe that the distance between two maximum points of AB solution is inversely proportional to the square root of the real value of the four wave mixing parameter $b$, so that when we increase the value of four wave mixing parameter the distance (\ref{dx2}) gets decreased and hence the number of the pulses is also increased. 
We illustrate the effect of $Re$ $b$ in Figs. 4.  Figs. 4(a) and 4(d) show the AB profile for the value $b=0.2+i$ which contain 4 peaks.  If we increase the real part of $b$ further, as happened in the GB case, the width of each profile shrinks and the distance between the consequent peaks decreases.  For example, when $b=1.5+i$ and $b=3.5+i$, the number of peaks in the AB profile increases to 6 and 8 peaks, respectively (for the same spatial distance and time interval), which is illustrated in Figs. 4.    
\par From physical point of view ABs have been realized experimentally in optical fibers \cite{abexp} specifically designed to study the breather evolution in a regime approaching the excitation of the Peregrine soliton.  Since four wave mixing effect is a fundamental feature of multi-mode fibers, the results presented here will be useful in the context of optical wave propagation in multi-mode fibers. 
\subsection{Ma soliton}
Restricting the imaginary part of the wave number to be zero, that is $\phi_R=0$ and $\phi_I\neq 0$ with $k=0$ in (\ref{pct5}) we get another breather solution which will propagate only in the time direction, that is
\begin{eqnarray}
p&=&\tau_1e^{i\theta}\nonumber\\&&\times\left(1+\frac{(\cosh(2\phi_I)-1)\cos\eta_I-i\sinh(2\phi_I)\sin\eta_I}{\sqrt{\vartheta}\cosh(\eta_R+\sigma)+\cos\eta_I}\right), \nonumber\\ 
q&=&\tau_2e^{i\theta}\nonumber\\&&\times\left(1+\frac{(\cosh(2\phi_I)-1)\cos\eta_I-i\sinh(2\phi_I)\sin\eta_I}{\sqrt{\vartheta}\cosh(\eta_R+\sigma)+\cos\eta_I}\right),\nonumber\\\label{pct8}
\end{eqnarray}
where $p_R$, $p_I$, $\Omega_R$, $\Omega_I$, $\eta_R$ and $\eta_I$ can be deduced from (\ref{pct5}) with $\phi_R=0$.
\begin{figure*}
\includegraphics[width=0.8\linewidth]{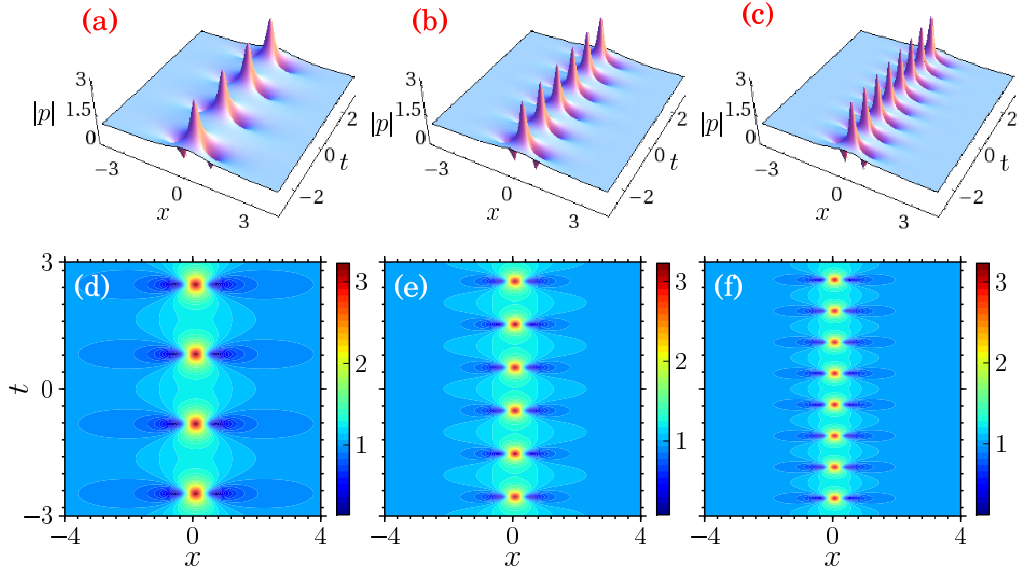}
\caption{(a) Ma breather profile of $p$ for the parameter values  $\tau_1 = 0.8$, $\tau_2=0.5$, $a=0.5$, $c=0.5$, $\phi_R = 0.7$, $\phi_I = -0.5$, $\eta_I^o=1.5$, $\eta_R^o=1.2$, $k=0$ with different four wave mixing parameters (a) $b=0.5+i$, (b) $b=1.5+i$, (c) $b=2.5+i$.  Figs. (d)-(f) are the corresponding contour plots of (a)-(c) respectively.  Similar profile occurs for $q$ also (not shown here)(Color online).}
\end{figure*}
We depict the solution (\ref{pct8}) in Figs. 5.  The plot confirms that the solution is periodic in $t$ and localized in $x$.  The wave solution which is temporally breathing and spatially oscillating is called a Ma breather/MS which was originally constructed by Ma for NLS equation \cite{Ma}.  We analyze the role of four wave mixing effect parameter $b$ analytically.  We find that a maximum point of Ma soliton solution occurs at 
\begin{eqnarray}
(x,t)=\left(\frac{-\sigma}{p_R},\frac{\pi}{\Omega_I}\right).
\end{eqnarray}
The next maximum point arrives at
\begin{eqnarray}
(x,t)=\left(\frac{-\sigma}{p_R},\frac{3\pi}{\Omega_I}\right).
\end{eqnarray}
The time duration between these two maximum points is
\begin{eqnarray}
d(t)=\frac{2\pi}{\Omega_I}=2\pi\frac{\cosh2\phi_I-\cos2\phi_R}{4Q(a\tau_1^2+c\tau_2^2+(b+b^*)\tau_1\tau_2)},
\label{dt2}
\end{eqnarray}
where
\begin{eqnarray}
Q&=&\sin2\phi_R(\cos^2\phi_R\sinh^2\phi_I-\sin^2\phi_R\cosh^2\phi_I)\nonumber\\&&-2\sinh2\phi_I(\cos\phi_R\sin\phi_R\cosh\phi_I\sinh\phi_I).
\end{eqnarray}
Hence from Eq. (\ref{dt2}), one can observe that the time duration between two maximum points is inversely proportional to the real value of the four wave mixing parameter $b$, so that when we increase the value of $Re$ $b$, the time duration between the maximum points decreases.  We illustrate the effect of $Re$ $b$ in Figs. 5.  Figs. 5(a) and 5(c) show the Ma soliton profiles for the $b$ value $b=0.5+i$, which contain 4 peaks.  When we increase the $b$ value to $b=1.5+i$ and $b=2.5+i$, the profile gets 6 and 8 peaks respectively which is illustrated in Figs. 5.  In other words the behavior replicates what has happened in the GB and AB cases. 
\par Recently Ma solitons have also been realized experimentally in optical fibers \cite{maexp}.  We believe that the Ma soliton solutions, (\ref{pct8}), of CGNLS equation (\ref{cnls01}) can also be realized in the context of wave propagation in the experimental multi-mode optical fibers, where four wave mixing becomes relevant.  
\subsection{RW solution}
\begin{figure*}
\includegraphics[width=0.8\linewidth]{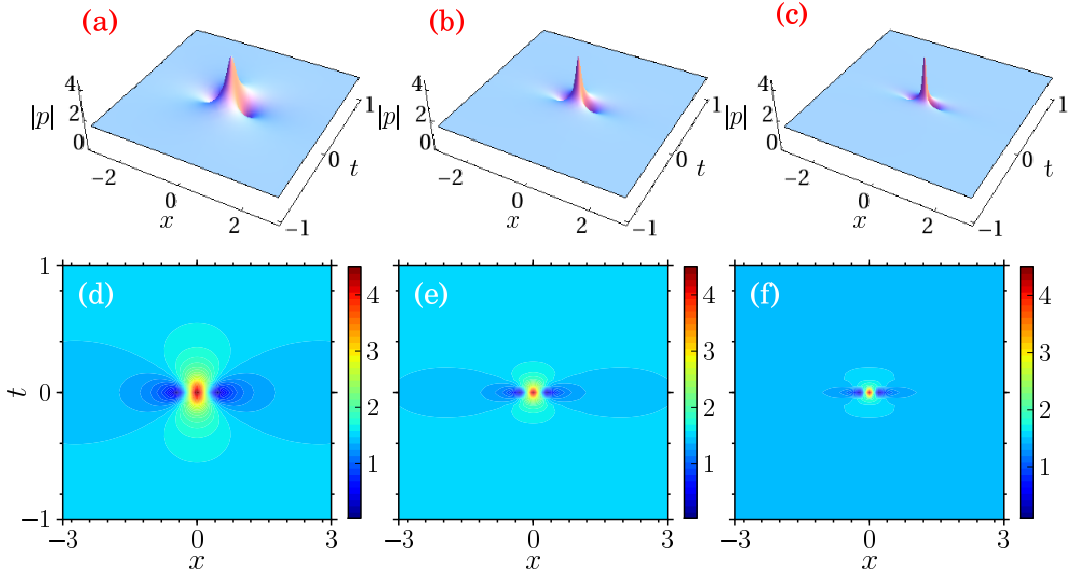}
\caption{(a) RW profile of $p$ for the parameter values $\tau_1=0.8$, $\tau_2=1.5$, $k=0$, $a=1$, $c=1$ with different four wave mixing parameter values (a)$b=0.5+i$, (b) $b=2+i$, (c) $b=5+i$.  Figs. (d)-(f) are the corresponding contour plots of (a)-(c) respectively.  Similar profile occurs for $q$ also (not shown here)(Color online).}
\end{figure*}
Finally we consider both $\phi_R$ and $\phi_I$ to be non-zero quantities such as $\phi_R=\epsilon\gamma$ and $\phi_I=\epsilon\rho$, where $\epsilon$ is a small parameter, $\gamma$ and $\rho$ are constants. Evaluating the expression (\ref{pct5}) in the limit $\epsilon\to 0$ by implementing a Taylor expansion with the restriction $\eta_2=\eta_1^*$, $\phi_2=\phi_1^*+\pi$, we find
\begin{eqnarray}
p_R&=&-2\sqrt{(a\tau_1^2+c\tau_2^2+(b+b^*)\tau_1\tau_2)}\rho\epsilon+O(\epsilon^3),\nonumber\\
p_I&=&2\sqrt{(a\tau_1^2+c\tau_2^2+(b+b^*)\tau_1\tau_2)}\gamma\epsilon+O(\epsilon^3),\nonumber\\
\Omega_R&=&(4(a\tau_1^2+c\tau_2^2+(b+b^*)\tau_1\tau_2)\gamma\nonumber\\&&-4k\sqrt{(a\tau_1^2+c\tau_2^2+(b+b^*)\tau_1\tau_2)}\rho)\epsilon+O(\epsilon^3),\nonumber\\
\Omega_I&=&(4(a\tau_1^2+c\tau_2^2+(b+b^*)\tau_1\tau_2)\rho\nonumber\\&&+4k\sqrt{(a\tau_1^2+c\tau_2^2+(b+b^*)\tau_1\tau_2)}\gamma)\epsilon+O(\epsilon^3),\nonumber\\
\sqrt{\vartheta}&=&1+\frac{1}{2}(\gamma^2+\rho^2)\epsilon^2,\nonumber\\
f&=&((\tilde{\eta}_R^2+\tilde{\eta}_I^2)+(\gamma^2+\rho^2))\epsilon^2+O(\epsilon^3),\nonumber\\
g&=&((\tilde{\eta}_R^2+\tilde{\eta}_I^2)-3(\gamma^2+\rho^2)+4i(\gamma \tilde{\eta}_R+\rho\tilde{\eta}_I))\epsilon^2\nonumber\\&&+O(\epsilon^3).\nonumber\\
\eta_R-\eta_R^0&=&\epsilon\tilde{\eta}_R+O(\epsilon^2)\nonumber\\
\eta_I-\eta_I^0&=&\epsilon\tilde{\eta_I}+O(\epsilon^2).
\label{pct10}
\end{eqnarray}
\par Substituting the above expressions (\ref{pct10}) into (\ref{pct5}) and taking the limit $\epsilon\to 0$ in the resultant expression, we find
\begin{eqnarray}
p&=&\tau_1e^{i\theta}\left(1-\frac{L}{M}\right),\;\;q=\tau_2e^{i\theta}\left(1-\frac{L}{M}\right),\nonumber\\
L&=&4+16i(a\tau_1^2+c\tau_2^2+(b+b^*)\tau_1\tau_2)t,\nonumber\\
M&=&1+4(a\tau_1^2+c\tau_2^2+(b+b^*)\tau_1\tau_2)(x-2kt)^2\nonumber\\&&+16(a\tau_1^2+c\tau_2^2+(b+b^*)\tau_1\tau_2)^2t^2,
\label{pct11}
\end{eqnarray} 
which is nothing but the RW solution of the CGNLS system.  The solution (\ref{pct11}) is localized both in space and time.  We note that the restriction $\tau_2=0$ and $b=0$ in (\ref{pct11}) provides the RW solution of the scalar NLS equation.  A typical evolution of the RW is shown in Figs. 6.  The amplitude of RW (\ref{pct11}) is
\begin{eqnarray}
|p|(x=0,t=0)=3\tau_1, \quad |q|(x=0,t=0)=3\tau_2.
\end{eqnarray}
\par It is interesting to analyze the role of the parameter $b$ in the RW solution as well.  Fig. 6(a) and 6(d) show the RW profile $b=0.5+i$.  When we increase the value of $Re$ $b$, the width of the profile gets shrunk.  Figs. 6(b) and  6(c) show the RW profile for $b=2+i$ and $b=5+i$, respectively, and Figs. 6(e) and 6(f) are their corresponding contour plots.  From these figures we can see that the RW becomes thinner and thinner when we increase the value of $Re$ $b$.  In other words the four wave mixing parameter $b$ plays the role of a perturbation parameter.  We can prove this behavior mathematically by finding the full width at half maximum (FWHM) of RW solution.  If $x_0$ is a half maximum point of RW, then at $t=0$,
\begin{eqnarray}
\tau_1\left(\frac{4(a\tau_1^2+c\tau_2^2+(b+b^*)\tau_1\tau_2)x_0^2-3}{1+4(a\tau_1^2+c\tau_2^2+(b+b^*)\tau_1\tau_2)x_0^2}\right)&=&\frac{1}{2}|p|_{max}\nonumber\\&=&\frac{1}{2}3\tau_1
\label{fwhm1}
\end{eqnarray}
From Eq. (\ref{fwhm1}), we get the half maximum points at
\begin{eqnarray}
x_0=\pm\frac{3}{2\sqrt{a\tau_1^2+c\tau_2^2+(b+b^*)\tau_1\tau_2}}.
\end{eqnarray}
The width between these two points is 
\begin{eqnarray}
FWHM=x_{0+}-x_{0-}=\frac{3}{\sqrt{a\tau_1^2+c\tau_2^2+(b+b^*)\tau_1\tau_2}}.
\label{fwhm2}
\end{eqnarray}  
Hence from Eq. (\ref{fwhm2}) we can see that the FWHM of RW is inversely proportional to the square root of the real value of the four wave mixing parameter $b$ so that when we increase the value of $Re$ $b$ the width of the pulse gets decreased.
\par As noted in the introduction, RWs are observed experimentally in the optical fibers, Bose-Einstein condensates, Plasma physics, super fluid He and so on.  Consequently the RW solution that exists in CGNLS system (\ref{cnls01}) with four wave mixing effects will be useful in the experimental realization of wave propagation in multi-mode fibers, multi-component plasmas and multi-component Bose-Einstein condensates.
\section{Akhmediev breather from RW solution}
\label{RWtoAB}
In Sec. III, we derived AB, MB and RW solutions from the GB solution.  In this section we construct the above solutions from the RW solution itself in a reverse way \cite{{Tajiri},{Vishnu}}.  To derive AB from RW solution we first factorize the RW solution (\ref{pct11}) in the following form, namely
\begin{eqnarray}
p&=&\tau_1\exp(i(kx-(k^2-2(a\tau_1^2+c\tau_2^2+(b+b^*)\tau_1\tau_2))t))\nonumber\\&&\times\left(1+\frac{1}{Q_1+R_1}\right)\left(1+\frac{1}{Q_1-R_1}\right),\nonumber\\
q&=&\tau_2\exp(i(kx-(k^2-2(a\tau_1^2+c\tau_2^2+(b+b^*)\tau_1\tau_2))t))\nonumber\\&&\times\left(1+\frac{1}{Q_1+R_1}\right)\left(1+\frac{1}{Q_1-R_1}\right),\nonumber\\
Q_1&=&2i(a\tau_1^2+c\tau_2^2+(b+b^*)\tau_1\tau_2)t,\nonumber\\
R_1&=&\frac{1}{2}\sqrt{1+4(a\tau_1^2+c\tau_2^2+(b+b^*)\tau_1\tau_2)(x-2kt)^2}.
\label{pct12}
\end{eqnarray}
For mathematical simplicity we take $k=0$.  With this choice Eq. (\ref{pct12}) can be rewritten in a more general form, that is
\begin{eqnarray}
p&=&\tau_1\exp(i(\sigma t+\phi))\left(1+g\sum_{n=-\infty}^{\infty}\frac{1}{i\alpha t+v(x)+n}\right)\nonumber\\&&\times\left(1+g\sum_{n=-\infty}^{\infty}\frac{1}{i\alpha t-v(x)+n}\right),
\nonumber\\
q&=&\tau_2\exp(i(\sigma t+\phi))\left(1+g\sum_{n=-\infty}^{\infty}\frac{1}{i\alpha t+v(x)+n}\right)\nonumber\\&&\times\left(1+g\sum_{n=-\infty}^{\infty}\frac{1}{i\alpha t-v(x)+n}\right),
\label{pct13}
\end{eqnarray}
where $g$ is a constant, $\alpha$, $\sigma$ and $v(x)$ are all to be determined.  We note that the spatial variable $x$ and the time variable $t$ are grouped with the real and imaginary parts respectively.  We have superposed the RW solutions in the $x$ direction. Using the trigonometric identity \cite{Integrals} $\cot\pi x=\frac{1}{\pi x}+\frac{x}{\pi}\sum_{n=-\infty}^\infty\frac{1}{n(x-n)}, \ n\neq 0$, we replace the infinite series by cot function and rewrite (\ref{pct13}) in a more compact form as
\begin{eqnarray}
p&=&\tau_1\exp(i(\sigma t+\phi))(1+g\pi \cot(\pi(v(x)+i\alpha t)))\nonumber\\&&\times(1-g\pi \cot(\pi(v(x)-i\alpha t))),\nonumber\\
q&=&\tau_2\exp(i(\sigma t+\phi))(1+g\pi \cot(\pi(v(x)+i\alpha t)))\nonumber\\&&\times(1-g\pi \cot(\pi(v(x)-i\alpha t))).
\label{pct14}
\end{eqnarray} 
\par Substituting the above expression (\ref{pct14}) in (\ref{cnls01}) and solving the resultant system of equations, we find
\begin{eqnarray}
\sigma&=&2(a\tau_1^2+c\tau_2^2+(b+b^*)\tau_1\tau_2)(1+\pi^2g^2)^2,
\label{pct18}\nonumber\\
\alpha&=&2(a\tau_1^2+c\tau_2^2+(b+b^*)\tau_1\tau_2)(1+\pi^2g^2)g,
\label{pct23}\nonumber\\
v(x)&=&\frac{1}{2\pi}\arccos\left(\frac{1}{\sqrt{1+\pi^2g^2}}\cos(\sqrt{2\pi^2\alpha g}x+v_0)\right),\nonumber\\\label{pct24}
\end{eqnarray}
where $v_0$ is a constant of integration.  Now substituting the expressions of $\sigma$, $\alpha$ and $v(x)$ in the general form (\ref{pct14}) and suitably rewriting the resultant expressions, we obtain the AB solution of the form
\begin{eqnarray}
p&=&\tau_1(1+\pi^2g^2)\left(1-\frac{E_1}{F_1}\right)\nonumber\\&\times&\exp(i(2(a\tau_1^2+c\tau_2^2+(b+b^*)\tau_1\tau_2)(1+\pi^2g^2)^2t+\phi)),\nonumber\\
q&=&\tau_2(1+\pi^2g^2)\left(1-\frac{E_1}{F_1}\right)\nonumber\\&\times&\exp(i(2(a\tau_1^2+c\tau_2^2+(b+b^*)\tau_1\tau_2)(1+\pi^2g^2)^2t+\phi)),\nonumber\\
E_1&=&(2\pi g)(\pi g\cosh2\pi\alpha t+i\sinh2\pi\alpha t),\nonumber\\
F_1&=&(1+\pi^2g^2)(\cosh2\pi\alpha t-(1/\sqrt{1+\pi^2g^2})\nonumber\\&&\times\cos(\sqrt{2\pi^2\alpha g}x+v_0)).
\label{pct25}
\end{eqnarray}
\par We note that the solution (\ref{pct25}) is periodic in the spatial direction and it grows exponentially fast in the initial stage from the time oscillatory background.  After reaching the maximum amplitude at a specific time, it decays exponentially again to the time oscillatory background \cite{Tajiri}.  A typical AB solution for a suitable set of parametric values is shown in Figs. 6.  
\begin{figure*}
\includegraphics[width=0.8\linewidth]{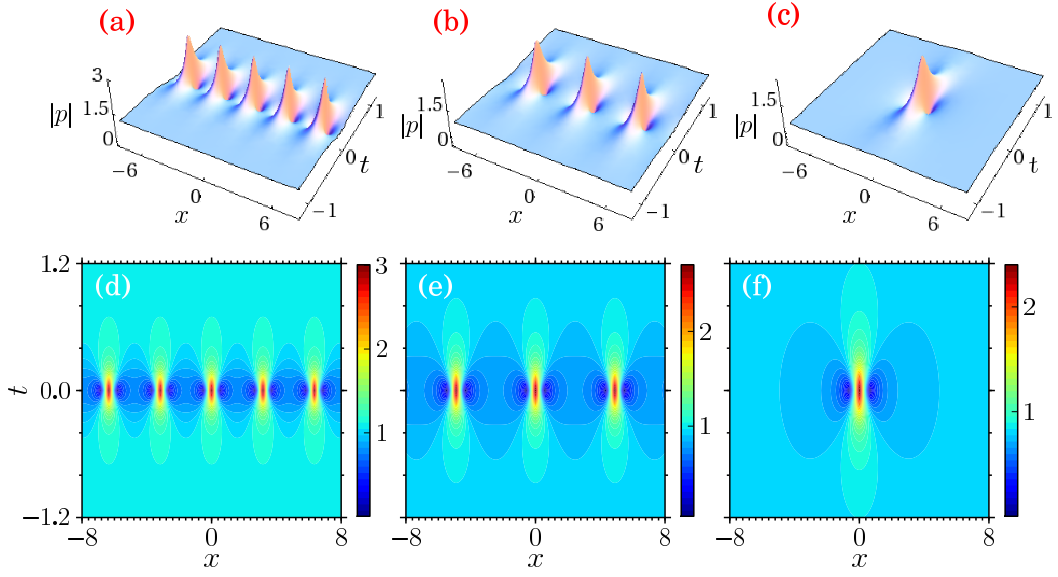}
\caption{(a) Akhmediev breather profile of $p$ for the parameter values $\tau_1=0.8$, $\tau_2=0.6$, $a=1$, $c=0.5$, $b=1+i$, $v_0=0$, $\phi=0$ with different critical parameter values (a) $g=0.2$, (b) $g=0.14$, (c) $g=0.02$.  Figs. (d)-(f) are the corresponding contour plots of (a)-(c) respectively.  Similar profile occurs for $q$ also (not shown here)(Color online).}
\end{figure*}
\par After carefully examining the AB profile we find that it strongly depends on the parameter $g$.  This  observation leads us to derive the RW from AB directly when the parameter $g$ goes to $0$.  If we decrease the $g$ value gradually to zero the number of peaks in AB profile decreases. Finally when $g$ attains a value which is nearly equal to zero we get a RW form.  To demonstrate this let us consider the AB profile in Fig. 7(a) which shows 5 peaks.  If we decrease the value of $g$, the temporal separation between adjacent peaks increases.  When $g=0.02$ it has only one peak.  Figs. 7(b) and 7(c) show this evolution clearly, Figs. 7(e)-7(f) are their corresponding contour plots.  Mathematically, by performing a Taylor expansion of the AB solution (\ref{pct25}) at $g=0$ one can obtain the RW solution whose expression exactly matches with the one given by (\ref{pct11}).  By comparing the AB solutions (\ref{pct7}) and (\ref{pct25}), one can observe that the parameter $g$ corresponds to the parameter $\phi_R$, since when $\phi_R\to 0$ in Eq. (\ref{pct7}), we can obtain the RW solution.
\par The method of obtaining RW from AB has been experimentally demonstrated in Ref. \cite{abexp} in the context of nonlinear optical fibers.  The critical parameter $g$ in our work corresponds to the modulation parameter `$a$' in \cite{abexp}.  So the results obtained in the present work can be useful for the experimental realization of them in birefringent optical fibers. 

\section{Ma soliton from RW solution}
\label{RWtoMS}
Now we focus our attention on deriving Ma soliton from RW solution.  To do so we rewrite the RW solution (\ref{pct11}) in a slightly different factorized form, that is
\begin{eqnarray}
p&=&\tau_1\exp(i(kx-(k^2-2(a\tau_1^2+c\tau_2^2+(b+b^*)\tau_1\tau_2))t))\nonumber\\
&&\times\left(1+\frac{i}{Q_2+R_2}\right)\left(1+\frac{i}{Q_2-R_2}\right),\nonumber\\
q&=&\tau_2\exp(i(kx-(k^2-2(a\tau_1^2+c\tau_2^2+(b+b^*)\tau_1\tau_2))t))\nonumber\\
&&\times\left(1+\frac{i}{Q_2+R_2}\right)\left(1+\frac{i}{Q_2-R_2}\right),\nonumber\\
Q_2&=&-2(a\tau_1^2+c\tau_2^2+(b+b^*)\tau_1\tau_2)t,\nonumber\\
R_2&=&\frac{i}{2}\sqrt{1+4(a\tau_1^2+c\tau_2^2+(b+b^*)\tau_1\tau_2)(x-2kt)^2}.
\label{pc12}
\end{eqnarray} 
and then consider it in a more general form with $k=0$, namely
\begin{eqnarray}
p&=&\tau_1\exp(i(\zeta t+\phi))\left(1+ih\sum_{n=-\infty}^{\infty}\frac{1}{\kappa t+i\varrho(x)+n}\right)\nonumber\\&&\times\left(1+ih\sum_{n=-\infty}^{\infty}\frac{1}{\kappa t-i\varrho(x)+n}\right),
\nonumber
\end{eqnarray}
\begin{eqnarray}
q&=&\tau_2\exp(i(\zeta t+\phi))\left(1+ih\sum_{n=-\infty}^{\infty}\frac{1}{\kappa t+i\varrho(x)+n}\right)\nonumber\\&&\times\left(1+ih\sum_{n=-\infty}^{\infty}\frac{1}{\kappa t-i\varrho(x)+n}\right),
\label{pct26}
\end{eqnarray}
with the assumption that the function $\varrho(x)$ and the parameters $\kappa$ and $\zeta$ are to be determined.  One may observe here that the RW is superposed in the temporal direction.  We identify the infinite series with the $\coth$ hyperbolic function \cite{Integrals}, $\coth\pi x=\frac{1}{\pi x}-\frac{ix}{\pi}\sum_{n=-\infty}^\infty\frac{1}{n(x-in)}, n\neq 0$, and rewrite the above expression as 
\begin{eqnarray}
p&=&\tau_1\exp(i(\zeta t+\phi))(1+h\pi \coth(\pi(\varrho(x)-i\kappa t)))\nonumber\\&&\times(1-h\pi \coth(\pi(\varrho(x)+i\kappa t))),\nonumber\\
q&=&\tau_2\exp(i(\zeta t+\phi))(1+h\pi \coth(\pi(\varrho(x)-i\kappa t)))\nonumber\\&&\times(1-h\pi \coth(\pi(\varrho(x)+i\kappa t))).
\label{pct27}
\end{eqnarray} 
\par Plugging the expressions (\ref{pct27}) into (\ref{cnls01}) and solving the resultant system of equations, after a very lengthy calculation, we find  
\begin{eqnarray}
\zeta&=&2(a\tau_1^2+c\tau_2^2+(b+b^*)\tau_1\tau_2)(1-\pi^2h^2)^2,\nonumber\\
\kappa&=&-2(a\tau_1^2+c\tau_2^2+(b+b^*)\tau_1\tau_2)(1-\pi^2h^2)h,\nonumber\\
\varrho(x)&=&\frac{1}{2\pi}\cosh^{-1}\bigg(\frac{1}{\sqrt{1-\pi^2h^2}}\cosh(\sqrt{-2\pi^2\kappa h}x+\varrho_0)\bigg),\nonumber\\
\label{pct35}
\end{eqnarray}
where $\varrho_0$ is a constant. 
\par  With these expressions, the general form of (\ref{pct26}) now becomes
\begin{eqnarray}
p&=&\tau_1(1-\pi^2h^2)\left(1+\frac{E_2}{F_2}\right)\nonumber\\&&\times\exp(i(2(a\tau_1^2+c\tau_2^2+(b+b^*)\tau_1\tau_2)(1-\pi^2h^2)^2t+\phi)),\nonumber\\
q&=&\tau_2(1-\pi^2h^2)\left(1+\frac{E_2}{F_2}\right)\nonumber\\&&\times\exp(i(2(a\tau_1^2+c\tau_2^2+(b+b^*)\tau_1\tau_2)(1-\pi^2h^2)^2t+\phi)),\nonumber\\
E_2&=&(2\pi h)(\pi h\cos2\pi\kappa t-i\sin2\pi\kappa t),\nonumber\\
F_2&=&(1-\pi^2h^2)(\cos2\pi\kappa t-(1/\sqrt{1-\pi^2h^2})\nonumber\\&&\times\cosh(\sqrt{-2\pi^2\kappa h}x+c))
\label{pct36}
\end{eqnarray}
which is nothing but the Ma breather solution.  This solution is periodic in the temporal direction and localized in space.  It grows and decays recurrently in time oscillate background as in the case of NLS equation \cite{Tajiri}.  The Ma breather solution of CGNLS equations for a set of parametric values is shown in Figs. 8.
\begin{figure*}
\includegraphics[width=0.8\linewidth]{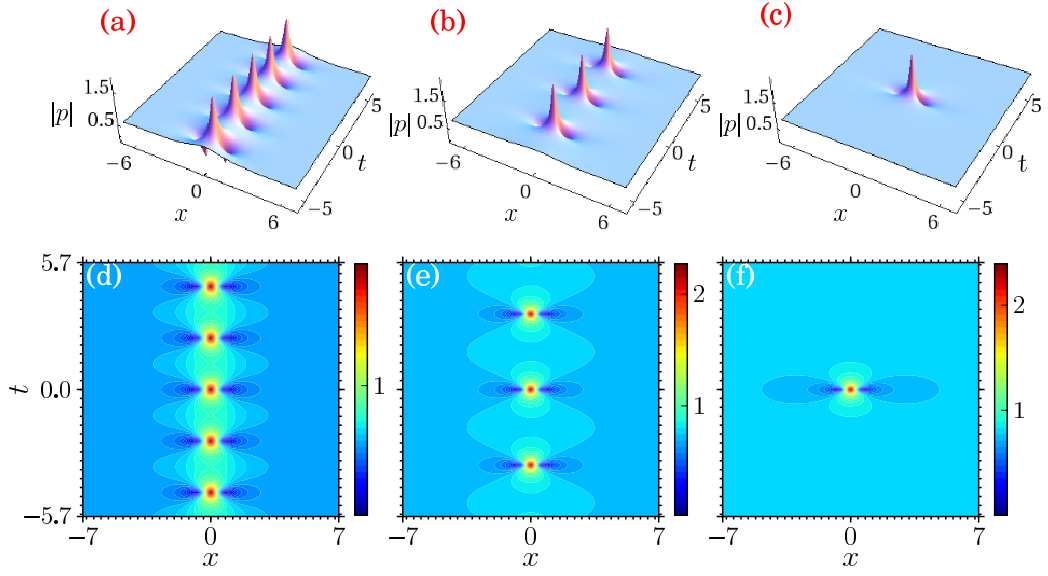}
\caption{(a) Ma breather profile of $p$ for the parameter values $\tau_1=0.8$, $\tau_2=0.6$, $a=1$, $c=1$, $b=1+i$  with different critical parameter values (a) $h=0.17$, (b) $h=0.09$, (c) $h=0.02$.  Figs. (d)-(f) are the corresponding contour plots of (a)-(c) respectively.  Similar profile occurs for $q$ also (not shown here)(Color online).}
\end{figure*}
\par The RW solution can also be obtained as the limiting case of Ma breathers.  This can be done by imposing the limit $h\to 0$ and incorporating the Taylor series expansion.  Here the critical parameter for getting RW is $h$.  If we decrease the $h$ value from $h=0.17$ to $h=0.02$ (Figs. 8) the distance between the peaks increases in the temporal direction.  At $h=0.02$ we get only a single peak (RW) even though we increase the time interval to (-10,000, 10,000).  The Taylor expansion at $h=0$ of (\ref{pct36}) yields the RW solution which matches with the one given in (\ref{pct11}).  By comparing the Ma breather solutions (\ref{pct8}) and (\ref{pct36}), one can check that the parameter $g$ corresponds to the parameter $\phi_I$, since when $\phi_I\to 0$ in Eq. (\ref{pct8}), we can get the RW solution. 

\section{General breather from RW} 
\label{RWtoGB}

Since GB solution is periodic in both $x$ and $t$ directions we need to include two complex arbitrary functions in the imbricate series of RWs, which in turn poses mathematical difficulties in transforming these imbricate series into trigonometric and hyperbolic functions.  So it will be very difficult to derive the GB solution from the RW solution in the same procedure as we did in the previous two cases. To overcome this difficulty we adopt the following methodology.  
\par To start with we rewrite the RW solution (\ref{pct11}) in the form $|p|^2=\tau_1^2(1-Q)(1-Q^*)$, $|q|^2=\tau_2^2(1-Q)(1-Q^*)$, with $Q$ as given in Eq. (\ref{pct11}) and the star denotes complex conjugate of it, and then reshape it as the second derivative of a logarithimic function, namely
\begin{eqnarray}
|p|^2&=&\tau_1^2-\frac{\tau_1^2}{2(a\tau_1^2+c\tau_2^2+(b+b^*)\tau_1\tau_2)}\nonumber\\&&\times\frac{\partial^2}{\partial x^2}\ln\bigg(\frac{1}{S} \times \frac{1}{T}\bigg),\nonumber\\
|q|^2&=&\tau_2^2-\frac{\tau_2^2}{2(a\tau_1^2+c\tau_2^2+(b+b^*)\tau_1\tau_2)}\nonumber\\&&\times\frac{\partial^2}{\partial x^2}\ln\bigg(\frac{1}{S} \times \frac{1}{T}\bigg),\label{pct37}
\end{eqnarray} 
where
\begin{eqnarray}
S&=&\bigg(\frac{1}{2}\sqrt{1+4(a\tau_1^2+c\tau_2^2+(b+b^*)\tau_1\tau_2)(x-2kt)^2}\nonumber\\&&+2i(a\tau_1^2+c\tau_2^2+(b+b^*)\tau_1\tau_2)t\bigg)^2,\nonumber\\
T&=&\bigg(\frac{1}{2}\sqrt{1+4(a\tau_1^2+c\tau_2^2+(b+b^*)\tau_1\tau_2)(x-2kt)^2}\nonumber\\&&-2i(a\tau_1^2+c\tau_2^2+(b+b^*)\tau_1\tau_2)t\bigg)^2.\nonumber
\end{eqnarray}
\par Now we consider Eq. (\ref{pct37}) in a more general form in order to superpose RWs in both space and time directions, that is
\begin{eqnarray}
|p|^2&=&\tau_1^2-\frac{\tau_1^2}{2(a\tau_1^2+c\tau_2^2+(b+b^*)\tau_1\tau_2)}\nonumber\\&&\times\frac{\partial^2}{\partial x^2}\ln\bigg(\sum_{n=-\infty}^\infty\frac{1}{(\phi(x,t)-i\psi(x,t)-n)^2}\nonumber\\&&\times\sum_{n=-\infty}^\infty\frac{1}{(\phi(x,t)+i\psi(x,t)-n)^2}\bigg),\nonumber\\
|q|^2&=&\tau_2^2-\frac{\tau_2^2}{2(a\tau_1^2+c\tau_2^2+(b+b^*)\tau_1\tau_2)}\nonumber\\&&\times\frac{\partial^2}{\partial x^2}\ln\bigg(\sum_{n=-\infty}^\infty\frac{1}{(\phi(x,t)-i\psi(x,t)-n)^2}\nonumber\\&&\times\sum_{n=-\infty}^\infty\frac{1}{(\phi(x,t)+i\psi(x,t)-n)^2}\bigg),\label{pct38}
\end{eqnarray}
where $\phi(x,t)$ and $\psi(x,t)$ are arbitrary functions of $x$ and $t$ which need to be determined.  Using the trigonometric identity \cite{Integrals} $\csc^2(\pi x)=\frac{1}{\pi^2}\sum_{n=-\infty}^\infty\frac{1}{x-n^2}$, the above expressions can be brought to the form,  
\begin{eqnarray}
|p|^2&=&\tau_1^2+\frac{\tau_1^2}{(a\tau_1^2+c\tau_2^2+(b+b^*)\tau_1\tau_2)}\nonumber\\&&\times\frac{\partial^2}{\partial x^2}\ln[\cosh2\pi\psi-\cos2\pi\phi],\nonumber\\
|q|^2&=&\tau_2^2+\frac{\tau_2^2}{(a\tau_1^2+c\tau_2^2+(b+b^*)\tau_1\tau_2)}\nonumber\\&&\times\frac{\partial^2}{\partial x^2}\ln[\cosh2\pi\psi-\cos2\pi\phi].\nonumber\label{pct40}
\end{eqnarray}
\par Instead of proceeding in the same direction and obtain GB solution we try to rewrite the GB solution which we already know to this problem.  To do so we consider the GB solution (\ref{pct5}) in the form,
\begin{eqnarray}
|p|^2=\tau_1^2+\frac{\tau_1^2}{(a\tau_1^2+c\tau_2^2+(b+b^*)\tau_1\tau_2)}\frac{\partial^2}{\partial x^2}\ln f,\nonumber\\
|q|^2=\tau_2^2+\frac{\tau_2^2}{(a\tau_1^2+c\tau_2^2+(b+b^*)\tau_1\tau_2)}\frac{\partial^2}{\partial x^2}\ln f, \label{pct41}
\end{eqnarray}
where $f=1+2e^{\eta_R}\cos\eta_I+ae^{2\eta_R}$, $\eta_R=p_Rx-\Omega_Rt+\eta_R^0$ and $\eta_I=p_Ix-\Omega_It+\eta_I^0$.  Comparing (\ref{pct41}) with the one derived from the RW solution we rewrite $f$ as $f=2e^{\eta_R}(\sqrt{\vartheta}\cosh(\eta_R+\sigma)$$-\cos(\eta_I+\theta))$ with $\eta_R$ and $\eta_I$ as given above.  The resultant expression now turns out to be
\begin{eqnarray}
|p|^2&=&\tau_1^2+\frac{\tau_1^2}{(a\tau_1^2+c\tau_2^2+(b+b^*)\tau_1\tau_2)}\nonumber\\&&\times\frac{\partial^2}{\partial x^2}\ln[\sqrt{\vartheta}\cosh(p_Rx-\Omega_Rt+\sigma)\nonumber\\&&-\cos(p_Ix-\Omega_It+\theta)],\nonumber\\
|q|^2&=&\tau_2^2+\frac{\tau_2^2}{(a\tau_1^2+c\tau_2^2+(b+b^*)\tau_1\tau_2)}\nonumber\\&&\times\frac{\partial^2}{\partial x^2}\ln[\sqrt{\vartheta}\cosh(p_Rx-\Omega_Rt+\sigma)\nonumber\\&&-\cos(p_Ix-\Omega_It+\theta)],\label{pct42}
\end{eqnarray}
where $\sigma=\eta_R^0+\frac{1}{2}\ln \vartheta$ and $\theta=\eta_I^0+\pi$.  
\par Now let us compare the two expressions $|p|^2$ and $|q|^2$, the one derived from the RW solution (vide Eq. (\ref{pct40})) and the other derived from the GB solution (vide Eq. (\ref{pct42})).  Doing so we find that both the expressions match in the following two cases, namely
\begin{eqnarray}
(i) \; \cosh 2\pi\psi&=&\sqrt{\vartheta}\cosh(p_Rx-\Omega_Rt+\sigma), \nonumber\\\cos 2\pi\phi&=&\cos(p_Ix-\Omega_It+\theta),\nonumber\\
(ii) \; \cosh 2\pi\psi&=&\cosh(p_Rx-\Omega_Rt+\sigma), \nonumber\\\cos 2\pi\phi&=&\frac{1}{\sqrt{\vartheta}}\cos(p_Ix-\Omega_It+\theta).
\end{eqnarray}
From these two sets of equations we find two different expressions for $\psi$ and $\phi$, namely
\begin{eqnarray}
(i)\;\psi&=&\frac{1}{2\pi}\ln(\sqrt{\vartheta}\cosh(p_Rx-\Omega_Rt+\sigma)\nonumber\\&&+\sqrt{\vartheta\cosh^2(p_Rx-\Omega_Rt+\sigma)-1}),\nonumber\\
\phi&=&\frac{1}{2\pi}(p_Ix-\Omega_It+\theta)\nonumber\\
(ii)\;\psi&=&\frac{1}{2\pi}(p_Rx-\Omega_Rt+\sigma),\nonumber\\ \phi&=&\frac{1}{2\pi}\arccos\left(\frac{1}{\sqrt{\vartheta}}\cos(p_Ix-\Omega_It+\theta)\right).\label{pct44}
\end{eqnarray}
Substituting (\ref{pct44}) into (\ref{pct38}) we obtain the imbricate series form of GB solution as
\begin{eqnarray}
|p|^2&=&\tau_1^2-\frac{\tau_1^2}{2(a\tau_1^2+c\tau_2^2+(b+b^*)\tau_1\tau_2)}\nonumber\\&&\times\frac{\partial^2}{\partial x^2}\ln\bigg(\sum_{n=-\infty}^\infty\frac{1}{(\frac{1}{2\pi}(p_Ix-\Omega_It+\theta)-J}\nonumber\\&&\times\sum_{n=-\infty}^\infty\frac{1}{(\frac{1}{2\pi}(p_Ix-\Omega_It+\theta)+J}\bigg),\nonumber\\
|q|^2&=&\tau_2^2-\frac{\tau_1^2}{2(a\tau_1^2+c\tau_2^2+(b+b^*)\tau_1\tau_2)}\nonumber\\&&\times\frac{\partial^2}{\partial x^2}\ln\bigg(\sum_{n=-\infty}^\infty\frac{1}{(\frac{1}{2\pi}(p_Ix-\Omega_It+\theta)-J}\nonumber\\&&\times\sum_{n=-\infty}^\infty\frac{1}{(\frac{1}{2\pi}(p_Ix-\Omega_It+\theta)+J}\bigg),\label{pct45}
\end{eqnarray}
where
\begin{eqnarray}
J&=&\frac{i}{2\pi}\ln(\sqrt{\vartheta}\cosh(p_Rx-\Omega_Rt+\sigma)\nonumber\\&&+\sqrt{\vartheta\cosh^2(p_Rx-\Omega_Rt+\sigma)-1})-n)^2.
\end{eqnarray}
The above solution is periodic in both space and time.  Since we have already presented the GB profile we do not discuss the characteristic properties of them here.

\section{Conclusion}
\label{conclusions} 
In this work, we have constructed the dark-dark soliton solution for the defocusing CGNLS system.  We have also derived a class of rational solutions, namely, GB, AB, MS and RW for the focusing CGNLS equation (\ref{cnls01}).  The Darboux transformation is widely used to derive breather and rational solutions of nonlinear Schr\"{o}dinger equation, derivative NLS type equations and their generalizations.  Differing from the conventional approach we have chosen Hirota's method to derive these solutions.  In particular we obtained the two soliton solution using Hirota's bilinearization method with a plane wave background.  From the two soliton solution we have deduced the explicit form of GB solution.  From the latter we have derived the AB, MS and RW solutions.  We have also analyzed how these solution profiles vary with respect to the four wave mixing parameter $b$.  Our results show that all the breather profiles strongly depend on value of $Re$ $b$.  While we increase the value of $Re$ $b$, the number of peaks in the breather profile increases and the width of each peak gets shrunk.  One notable behavior in GB which we have observed is that the direction of this profile also changes when we increase the value of $Re$ $b$.  As far as the RW profile is concerned the width of the peak becomes very thin when we increase the value of $b$.  
\par In the second part of our work  we have derived AB, MS and GB from the RW solution as the starting point.  The expressions obtained in both the directions match with each other.  In the course of this analysis we have also demonstrated how to construct RW from AB and MS.  The AB and MS solutions depend on certain critical parameters.  When one of these parameters goes to zero we are able to capture the RW solution.  We also wish to add that we have also attempted to derive bright-dark soliton solution of (\ref{cnls01}) for different choices of $c$ but we could not succeed.  It is unclear to us how to construct a bright-dark soliton solution for this equation, which we think is a problem of considerable interest for further investigation.  Finally, we also remark that the various solutions discussed in this paper, such as AB, MS and RW have been realized in the context of scalar NLS type evolution equations representing wave propagation in optical fibers, Bose-Einstein condensates, plasma physics, oceanography and so on.  We expect that our results on multicomponent CGNLS system can lead to the realization of the above type of coherent structures in multi-mode systems where four wave mixing is important.      

\section*{Acknowledgements}
NVP wishes to thank the University Grants Commission (UGC-RFSMS), 
Government of India, for providing a Research Fellowship. The work of MS forms part of a research project sponsored by NBHM, Government of India and while the work of ML forms part of an IRHPA project and a Ramanna Fellowship project 
of ML, sponsored by the Department of Science and Technology (DST), Government of India. ML 
also acknowledges the financial support under a DAE Raja Ramanna Fellowship. 

\appendix
\section{Derivatives of GB solution}
Differentiating GB solution (\ref{pct5}) we obtain the following expressions.
\begin{eqnarray}
|p|_x&=&\frac{\tau_1}{2\sqrt{\frac{A^2+B^2}{D^2}}}\bigg[\frac{2A}{D^2}(-p_I\cosh2\phi_I\sin\eta_I+\sqrt{a}p_R\nonumber\\&&\times\cos2\phi_R\sinh(\eta_R+\sigma))+\frac{2B}{D^2}(\sqrt{a}p_R\cosh(\eta_R+\sigma)\nonumber\\&&\times\sin2\phi_R-p_I\cos\eta_I\sinh2\phi_I)-\frac{2}{D^3}(-p_I\nonumber\\&&\times\sin\eta_I+\sqrt{a}p_R\sinh(\eta_R+\sigma))(A^2+B^2)\bigg],
\end{eqnarray}
\begin{eqnarray}
|p|_t&=&\frac{\tau_1}{2\sqrt{\frac{A^2+B^2}{D^2}}}\bigg[\frac{2A}{D^2}(\Omega_I\cosh2\phi_I\sin\eta_I-\sqrt{a}\Omega_R\nonumber\\&&\times\cos2\phi_R\sinh(\eta_R+\sigma))+\frac{2B}{D^2}(-\sqrt{a}\Omega_R\nonumber\\&&\times\cosh(\eta_R+\sigma)\sin2\phi_R+\Omega_I\cos\eta_I\sinh2\phi_I)\nonumber\\&&-\frac{2}{D^3}(\Omega_I\sin\eta_I-\sqrt{a}\Omega_R\sinh(\eta_R+\sigma))(A^2+B^2)\bigg],\nonumber\\
\end{eqnarray}
\begin{eqnarray}
|q|_x&=&\frac{\tau_2}{2\sqrt{\frac{A^2+B^2}{D^2}}}\bigg[\frac{2A}{D^2}(-p_I\cosh2\phi_I\sin\eta_I+\sqrt{a}p_R\nonumber\\&&\times\cos2\phi_R\sinh(\eta_R+\sigma))+\frac{2B}{D^2}(\sqrt{a}p_R\cosh(\eta_R+\sigma)\nonumber\\&&\times\sin2\phi_R-p_I\cos\eta_I\sinh2\phi_I)-\frac{2}{D^3}(-p_I\nonumber\\&&\times\sin\eta_I+\sqrt{a}p_R\sinh(\eta_R+\sigma))(A^2+B^2)\bigg],
\end{eqnarray}
\begin{eqnarray}
|q|_t&=&\frac{\tau_1}{2\sqrt{\frac{A^2+B^2}{D^2}}}\bigg[\frac{2A}{D^2}(\Omega_I\cosh2\phi_I\sin\eta_I-\sqrt{a}\Omega_R\nonumber\\&&\times\cos2\phi_R\sinh(\eta_R+\sigma))+\frac{2B}{D^2}(-\sqrt{a}\Omega_R\nonumber\\&&\times\cosh(\eta_R+\sigma)\sin2\phi_R+\Omega_I\cos\eta_I\sinh2\phi_I)\nonumber\\&&-\frac{2}{D^3}(\Omega_I\sin\eta_I-\sqrt{a}\Omega_R\sinh(\eta_R+\sigma))(A^2+B^2)\bigg],\nonumber\\
\end{eqnarray}
where
\begin{eqnarray}
A&=&\cos\eta_I\cosh2\phi_I+\sqrt{a}\cos2\phi_R\cosh(\eta_R+\sigma),\nonumber\\
B&=&-\sin\eta_I\sinh2\phi_I+\sqrt{a}\sin2\phi_R\sinh(\eta_R+\sigma),\nonumber\\
D&=&\cos\eta_I+\sqrt{a}\cosh(\eta_R+\sigma).
\end{eqnarray}

\end{document}